\documentclass[aps,prl,superscriptaddress,reprint, onecolumn, floatfix]{revtex4-2}
\usepackage{graphicx}
\usepackage{siunitx}
\usepackage{upgreek} % µ in Roman in math environment: \upmu
\usepackage[colorlinks,citecolor=blue,urlcolor=blue,bookmarks=false,hypertexnames=true]{hyperref}

\begin{document}

\title{Control of the Bose--Einstein Condensation of Magnons by the Spin-Hall Effect}%
\author{Michael Schneider}
\email[e-Mail: ]{mi\_schne@rhrk.uni-kl.de}
\affiliation{Fachbereich Physik and Landesforschungszentrum OPTIMAS, Technische Universit\"at Kaiserslautern, D-67663 Kaiserslautern, Germany}
\author{David Breitbach}
\affiliation{Fachbereich Physik and Landesforschungszentrum OPTIMAS, Technische Universit\"at Kaiserslautern, D-67663 Kaiserslautern, Germany}
\author{Rostyslav O. Serha}
\affiliation{Fachbereich Physik and Landesforschungszentrum OPTIMAS, Technische Universit\"at Kaiserslautern, D-67663 Kaiserslautern, Germany}
\author{Qi Wang}
\affiliation{Faculty of Physics, University of Vienna, A-1090 Vienna, Austria}
\author{Alexander A. Serga}
\affiliation{Fachbereich Physik and Landesforschungszentrum OPTIMAS, Technische Universit\"at Kaiserslautern, D-67663 Kaiserslautern, Germany}
\author{Andrei N. Slavin}
\affiliation{Department of Physics, Oakland University, Rochester, MI, USA}
\author{Vasyl S. Tiberkevich}
\affiliation{Department of Physics, Oakland University, Rochester, MI, USA}
\author{Bj\"orn Heinz}
\affiliation{Fachbereich Physik and Landesforschungszentrum OPTIMAS, Technische Universit\"at Kaiserslautern, D-67663 Kaiserslautern, Germany}
\author{Bert L\"agel}
\affiliation{Fachbereich Physik and Landesforschungszentrum OPTIMAS, Technische Universit\"at Kaiserslautern, D-67663 Kaiserslautern, Germany}
\author{Thomas Br\"acher}
\affiliation{Fachbereich Physik and Landesforschungszentrum OPTIMAS, Technische Universit\"at Kaiserslautern, D-67663 Kaiserslautern, Germany}
\author{Carsten Dubs}
\affiliation{INNOVENT e.V. Technologieentwicklung, D-07745  Jena, Germany}
\author{Sebastian Knauer}
\affiliation{Faculty of Physics, University of Vienna, A-1090 Vienna, Austria}
\author{Oleksandr V. Dobrovolskiy}
\affiliation{Faculty of Physics, University of Vienna, A-1090 Vienna, Austria}
\author{Philipp Pirro}
\affiliation{Fachbereich Physik and Landesforschungszentrum OPTIMAS, Technische Universit\"at Kaiserslautern, D-67663 Kaiserslautern, Germany}
\author{Burkard Hillebrands}
\affiliation{Fachbereich Physik and Landesforschungszentrum OPTIMAS, Technische Universit\"at Kaiserslautern, D-67663 Kaiserslautern, Germany}
\author{Andrii V. Chumak}
\affiliation{Faculty of Physics, University of Vienna, A-1090 Vienna, Austria}%

\date{September 2021}%

\begin{abstract}
Previously, it has been shown that rapid cooling of yttrium-iron-garnet (YIG)/platinum (Pt) nano structures, preheated by an electric current sent through the Pt layer, leads to overpopulation of a magnon gas and to subsequent formation of a Bose--Einstein condensate (BEC) of magnons. The spin Hall effect (SHE), which  creates a spin-polarized current in the Pt layer, can inject or annihilate magnons depending on the electric current and applied field orientations. Here we demonstrate that the injection or annihilation of magnons via the SHE can prevent or promote the formation of a rapid cooling induced magnon BEC. Depending on the current polarity, a change in the BEC threshold of -8\% and +6\% was detected. These findings demonstrate a new method to control macroscopic quantum states, paving the way for their application in spintronic devices.
 \end{abstract}
 
\maketitle
%\tableofcontents
The Bose--Einstein condensate (BEC) \cite{einstein1924quantentheorie}, often considered because of its exotic properties as the fifth state of matter, is formed when individual atoms \cite{Anderson.1995}, subatomic particles \cite{Begun2008}, or quasiparticles such as Cooper pairs \cite{Cooper1956} or quanta of molecular electric oscillations \cite{Frohlich1968} coalesce into a single quantum mechanical entity existing on a macroscopic scale and described by a single wave function.
Emerging in various physical systems from neutron stars \cite{Danila2015} to a droplet of liquid helium \cite{Volovik.2009}, this state leads to fascinating and valuable macroscopic quantum phenomena, such as superconductivity and superfluidity.
Recently, novel BEC applications have been proposed, including those in the rapidly developing field of quantum computing \cite{nakata2014persistent_current, tserkovnyak2017, byrnes2012, adrianov2014, Xue2019polartiton_qubit, Ghosh2020exciton-polariton_quantum-comp}. In contrast to existing quantum computers, which operate at about 20\,$\mathrm{\upmu K}$ \cite{arute2019IBM}, BEC-based computing can be performed at much higher temperatures: for example, in yttrium iron garnet (Y$_3$Fe$_5$O$_{12}$, YIG) \cite{Cherepanov.1993}, a magnon condensate \cite{Demokritov.2006} was observed at room temperature. When using such a magnon condensate in both quasi-quantum and classical nanoscale devices, the possibility to control it by magnon spintronics \cite{Chumak.2015} methods via spin-polarized electric currents \cite{Safranski.2017} seems particularly attractive for reducing power consumption and simplifying these devices.

The formation of a Bose--Einstein condensate (BEC) can be achieved  by a decrease in the temperature for real-particle systems \cite{Anderson.1995,Davis1995}, or in a quasi-particle system by the injection of bosons resulting in an increase in the chemical potential. The latter has been demonstrated experimentally for exciton-polaritons \cite{Kasprzak.2006, Lerario.2017}, photons \cite{Klaers.2010, Damm.2016} or magnons  \cite{Bunkov2008, Nikuni.2000, Yin.2008, Giamarchi.2008,Demokritov.2006, Serga.2014, Bozhko.2016, Rezende.2009}. In the case of magnonic systems the injection was realised by the nuclear magnetic resonance \cite{Bunkov2008, Autti2020}, by the parametric pumping mechanism \cite{Demokritov.2006,Serga.2014,Bozhko.2016,Lvov.1994}, allowing for the injection of a large number of magnons at a given frequency, via the spin-Seebeck effect \cite{Safranski.2017} or by the mechanism of rapid cooling, as it was shown recently \cite{Schneider.2020}. The method of rapid cooling makes use of  the application of a DC heating pulse and the subsequent rapid cooling of yttrium iron garnet (YIG)/Pt nano structures. The heating generates a high population of magnons being in thermal equilibrium with the phononic system. A rapid decrease in the phononic temperature results in the break of the equilibrium. Since the lifetime of magnons is larger than the phonon cooling rate in the experiment  an overpopulation of magnons over the whole magnon spectrum is generated. This overpopulation results in a redistribution of magnons from higher to lower energies. In this way, if the temperature of the heated YIG film is high enough and the cooling process is fast enough, the magnon chemical potential is increased to the minimal energy of the magnon system and the BEC formation process is triggered. \looseness=-1

In the previous experiments, a Pt or Al layer was used to heat the YIG nano structure \cite{Schneider.2020}. For a Pt heater an additional generation of a spin-polarized current transverse to the YIG/Pt interface due to the spin Hall effect (SHE) is expected \cite{Collet.2016,Cornelissen.2016,Cornelissen.2018, Madami.2011}. The resulting spin current is known to act on the magnetization dynamics in the YIG via the spin transfer torque (STT) \cite{Demidov.2011,Demidov.2017, MichaelSchreier.2014, Chumak.2015, Saitoh.2006, Wimmer, PhysRevLett.113.197203}. The SHE-STT contribution, which can be easily checked by the variation of the current polarity with respect to the magnetization orientation in the YIG film \cite{MichaelSchreier.2014}, was not observed in the original experiments \cite{Schneider.2020}. The reason was the large YIG film thickness of 70\,nm (STT is an interface effect) and the high quality of the Pt layer grown by molecular beam epitaxy that results in a small spin Hall angle \cite{Sagasta.2016}. \looseness=-1

Here, we investigate a similar structure but with a smaller YIG film thickness of 34\,nm \cite{Dubs2020} and with the Pt layer deposited by a sputtering technique to achieve a pronounced SHE-STT effect. 
Using such YIG nano structures, we are now able to investigate the influence of the SHE-STT effect on the formation of the magnon BEC by rapid cooling. Analogously to our original studies, the experiments are conducted at room temperature. We show that the magnons annihilated or injected by the STT effect during the pulse application strongly influence the BEC formation threshold.

\begin{figure}[]
	\includegraphics[width=0.45\textwidth]{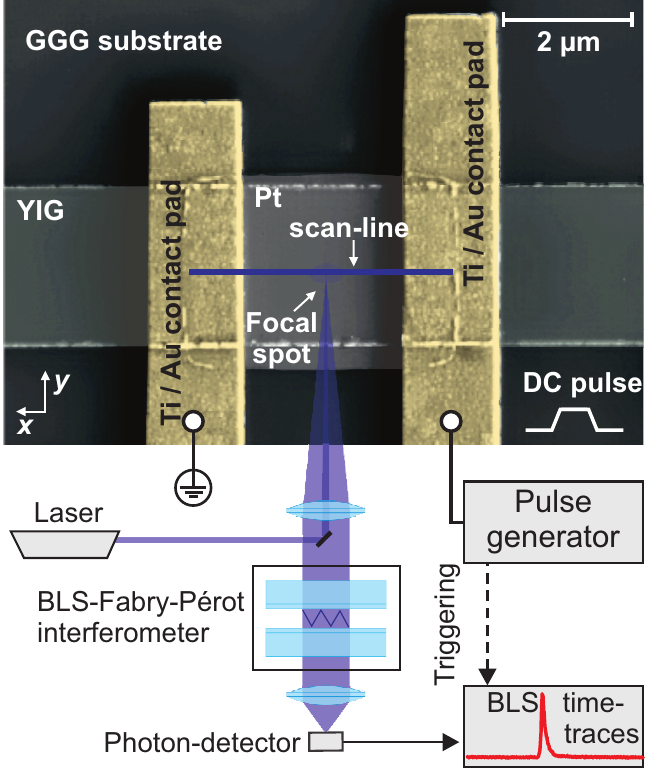}
	\caption{Colored scanning-electron microscopy-image of the structure under investigation and sketch of the experimental setup. The structure consists of a 2\,$\mathrm{\upmu m}$-wide and 34\,nm-thick YIG stripe. A 3\,$\mathrm{\upmu m}$-long and 7\,nm-thick platinum-heater is placed on top and contacted by Ti/Au-leads separated by a distance of 2\,$\mathrm{\upmu m}$. }~\label{figure1}
\end{figure}

Figure~\ref{figure1} shows the structure under investigation and a sketch of the experimental setup. The structure consists of a 2\,$\mathrm{\upmu m}$-wide YIG/Pt stripe ($\SI{34}{\nano\meter}$/$\SI{7}{\nano\meter}$) on a (111) gadolinium gallium garnet (GGG) substrate. 
The YIG structure was fabricated using electron-beam lithography with subsequent argon-ion milling \cite{Heinz.2020}. Afterwards a 3\,$\mathrm{\upmu m}$-long Pt layer was deposited on the waveguide using a RF-sputtering technique. To establish electrical contacts to the platinum pads, Ti/Au-leads (10\,nm/150\,nm) with a distance of 2\,$\mathrm{\upmu m}$ between the inner edges were fabricated by electron beam evaporation. In the presented experiments DC pulses of a duration of $\tau_\mathrm{P}=\SI{100}{ns}$ are applied to the Pt pad. Standard ferromagnetic resonance (FMR)  measurements were performed on two reference pads on the same sample, one bare and one covered with platinum. These yielded Gilbert-damping-constants of $\alpha_\mathrm{YIG} = 1.8\times 10^{-4}$ for the bare YIG pad and $\alpha_\mathrm{YIG|Pt} = 1.7\times 10^{-3}$ for the YIG/Pt pad, corresponding to a spin mixing conductance of $g^{\uparrow\downarrow}=\SI{5.1E+18}{\per\square\meter}$ \cite{Du.2015}. An external field of $\upmu_0 \mathbf{H}_\mathrm{ext}=\SI{110}{\milli\tesla}$ magnetizes the stripe either along its short or long axis in plane. We apply DC-current pulses to the Pt-layer in order to trigger the SHE-STT effect based injection or annihilation of magnons and to heat up the structure. After pulse termination, the structure cools down rapidly since the GGG substrate and the Au leads act as efficient heat sinks. In such a way, the experiments are conducted at room temperature and no active cooling is required. \color{black}
The magnetization dynamics is measured by means of Brillouin light scattering spectroscopy (BLS). A laser beam with $\SI{457}{\nano\meter}$ wavelength and $5.0 \pm 0.5$\,mW power is focused (spot size 400 nm) onto the YIG waveguide through the backside of the transparent GGG substrate, allowing to measure below the platinum-covered YIG region \cite{Heinz.2020}. The inelastically scattered light, carrying the information about the magnon intensity and frequency, is analyzed by a multi-pass tandem Fabry-P\'{e}rot interferometer with a time resolution of about $\SI{2}{\nano \second}$ \cite{Sebastian.2015, Schneider.2020}. The laser focus was scanned along the platinum layer between the two contact pads on a line in the middle of the stripe (see scan line indicated in Fig.\,\ref{figure1}). The resulting BLS signal was integrated along this scan line to reduce any influence of an inhomogeneous heating of the platinum region on the experimental results.

\begin{figure*}
		\includegraphics[width=1.0\textwidth]{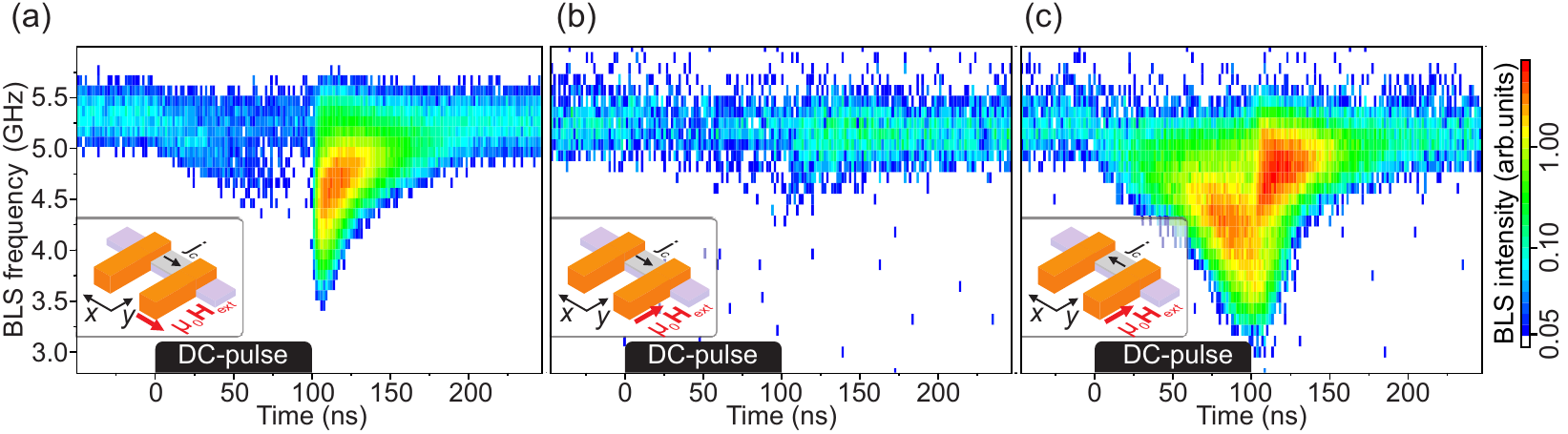}
		\caption{BLS intensity color-coded (log-scale) as a function of the BLS frequency and  time. The duration of the $100$-ns long heating DC pulse with an amplitude of $|U|=\SI{1.5}{V}$ is marked by the black boxes. (a) Current parallel to the external field, thus no contribution of the SHE-STT effect is expected. The achieved overpopulation after pulse termination due to the rapid cooling effect is sufficiently large to trigger the formation of a magnon BEC. (b) Current is perpendicular to the external field. The STT annihilates magnons during the pulse. The condensation of magnons is suppressed. (c) Reversed current direction in comparison to the situation in (b), the STT injects magnons. After the pulse is applied the magnon BEC density is enhanced compared to the situation depicted in panel (a). For the exact geometries see insets.}
		\label{figure_2}
	\end{figure*}

Figures\,\ref{figure_2}(a-c) show the measured color-coded BLS intensities (log-scale) as a function of time and the BLS frequency. The amplitude of the applied DC heating pulse is 
$|U|=\SI{1.5}{\volt}$, corresponding to a current density of $|j_\mathrm{C}|=\SI{1.6E+12}{\ampere\per\square\meter}$. 

Figure~\ref{figure_2}(a) shows the reference experiment with an external field $\upmu_0 \mathbf{H}$ aligned parallel to the long axis of the waveguide, parallel to the direction of the current $\mathbf{j}_\mathrm{C}$ ($\upmu_0 \mathbf{H}_\mathrm{ext}\parallel \mathbf{j}_\mathrm{C}$). In this geometry, no contribution of the SHE-STT effect is expected and the application of the current pulse only results in a Joule heating-induced increase of the YIG/Pt temperature. While the DC pulse is applied (black box in the figure), the BLS intensity decreases, as originally investigated in Ref.~\cite{Olsson.2017}. Further, the spectrum of thermal magnons shifts to lower frequencies due to the decrease in the saturation magnetization \cite{Cherepanov.1993, Schneider.2020}. After the DC pulse is switched off at $t = \SI{100}{\nano\second}$, the heat-dissipation-induced rapid cooling triggers the formation of the BEC at the bottom of the spectrum \cite{Schneider.2020}. The BEC manifests itself as a pronounced peak in the magnon intensity. The accompanying frequency increase is due to the cooling.

The contribution of the SHE can be switched on by rotating the magnetic field by $90^\circ$, hence pointing along the short axis of the stripe [see insets in Figs.\,\ref{figure_2}(b,c)], i.e. perpendicular to the direction of the DC current \cite{MichaelSchreier.2014}.

 In this geometry the SHE-STT contribution can be damping- or anti-damping like \cite{Collet.2016, V.Lauer.2017}. The change in the effective damping can also be described by the annihilation or injection of magnons \cite{MichaelSchreier.2014}, or by the change in the magnon chemical potential $\upmu$ \cite{Demidov.2017, Cornelissen.2016}. In the case without a STT-contribution  [Fig.~\ref{figure_2}(a), pulse duration of $\tau_\mathrm{P}=\SI{100}{ns}$ and a voltage of $U= \SI{1.5}{\volt}$)], the magnon and the phonon system are in thermal equilibrium at the end of the DC heating pulse, and both are highly populated.
  The SHE-STT contribution increases or decreases the number of magnons at the end of the pulse with respect to the reference case.
  Thus, the SHE-STT is expected to change the BEC formation via a change of the number and distribution of excess magnons prior to the rapid cooling process \cite{Schneider.2020}.

Figures\,\ref{figure_2}(b), and \ref{figure_2}(c) depict the cases for magnon annihilation and injection via the SHE-STT effect, respectively. These processes can be seen in the increase or decrease of BLS intensity during the DC pulse.

The magnon annihilation process [Fig.~\ref{figure_2}(b)], in contrast to experiments on thicker YIG micro structures used previously \cite{Schneider.2020}, is now large enough to compensate for the rapid cooling-induced increase of the chemical potential $\upmu$, thus suppressing BEC formation.

In the case of the SHE-STT-induced magnon injection process [Fig.~\ref{figure_2}(c), $j_{\mathrm{C},x}<0, \upmu_0 \mathbf{H}_\mathrm{ext}\perp \mathbf{j}_\mathrm{C}$] the SHE-STT effect enhances the magnon redistribution, which is manifested by the even higher BLS intensity measured after the DC pulse, compared to Fig.~\ref{figure_2}(a).
For a better comparison of the three cases described above, see Supplemental Material for extracted BLS spectra during and after the pulse.

Note that a purely thermal excitation of magnons takes place over the whole spectral range [Fig.~\ref{figure_2}(a)]. The subsequent decrease of the saturation magnetization leads to a decrease in the BLS intensity \cite{Olsson.2017}. In contrast, for a STT-injection of magnons the BLS intensity increases [Fig.\,\ref{figure_2}(c)], in spite of the heating-induced decrease of the BLS sensitivity (which does not depend on the field orientation). The reduction of the BLS sensitivity is given by the decrease in the saturation magnetization that can be treated as a thermally-induced increase of the number of magnons over the whole magnon spectrum \cite{WWettling.1975}. Thus, the fact that the BLS intensity in Fig.\,\ref{figure_2}(c) increases rather than decreases during the heating process suggests that the SHE-STT mechanism injects  magnons primarily into the low energy part of the spectrum accessible with micro focused BLS.

 	\begin{figure*}
\includegraphics[width=1.0\textwidth]{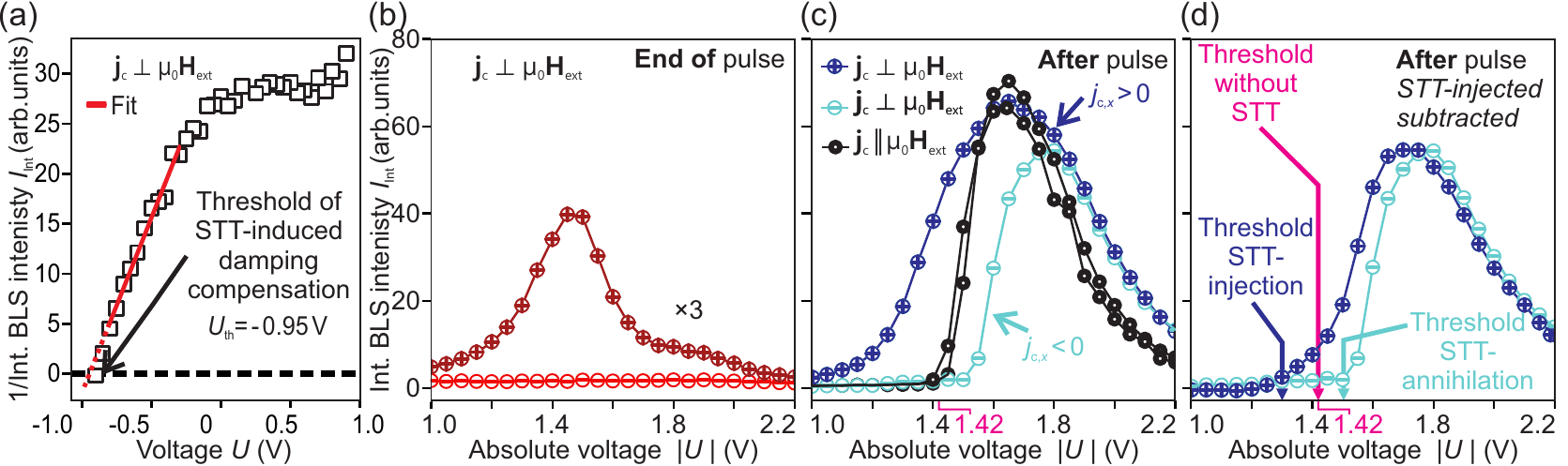}
\caption{(a) Inverse BLS intensity as a function of the voltage for the application of a continuous DC current. The linear fit yields a threshold for the damping compensation of $U_\mathrm{th}=\SI{-0.95}{\volt}$. (b) Integrated BLS intensity at the end of the applied DC pulse (integration interval $\tau_\mathrm{P}-\SI{8}{\nano \second}<t<\tau_\mathrm{P}$) as a function of the absolute value of the voltage. The external field of $\upmu_0 \mathbf{H}_\mathrm{ext}=\SI{110}{\milli\tesla}$ is aligned perpendicular to the direction of the current. The SHE-STT effect either injects (``$+$''-sign data points) or annihilates (``$-$''-sign data points) magnons. (c) Integrated BLS intensity after the pulse is switched off ($\tau_\mathrm{P}<t<\tau_\mathrm{P}+\SI{95}{\nano\second}$) as a function of the absolute value of the voltage. The blue curves correspond to the situation in (b), the black curves show the reference curve for $\upmu_0 \mathbf{H}_\mathrm{ext}\parallel \mathbf{j}_\mathrm{C}$, without a SHE-STT contribution. (d) Integrated intensity as in (c), the residual signal caused by the decaying STT-injected magnons is subtracted. Due to the injection or annihilation of magnons via the STT the thresholds are shifted with respect to the case without a SHE-STT contribution (pink line).}~\label{figure3}
\end{figure*}

The results presented in Fig.\,\ref{figure_2} show that the SHE-STT effect can control the BEC formation process via the injection or annihilation of magnons. Here we would like to point out that the measurements presented in Fig.~\ref{figure_2}(b,c) are conducted for a fixed geometry of the field. Thus, the difference in the BEC formation after pulse termination is solely determined by the SHE-STT effect. To investigate the threshold of the BEC formation process in the presence of the SHE-STT effect, we first characterize the  state of the magnon system right before the cooling takes place.

Figure~\ref{figure3}(a) shows the inverse BLS intensity as a function of the applied voltage, now for an applied continuous DC voltage instead of a DC pulse.  The linear fit of the inverse BLS intensity yields the threshold voltage for the STT-induced damping compensation \cite{Demidov.2011}, which is found to be $U=\SI{-0.95}{V}$. Thus, for the voltages applied in the pulsed experiments, the SHE-STT effect compensates the damping and should lead to the formation of auto-oscillations after a sufficiently long time \cite{V.Lauer.2017}. In the following, we limit our discussion to the voltage regime, where no auto-oscillations are triggered. However, we discuss measurements for higher voltages in the Supplemental materials.

Figure~\ref{figure3}(b) shows the BLS intensity integrated over the whole frequency range shown in Fig.\,\ref{figure_2} and integrated over the last \SI{8}{\nano\second} of the applied DC pulse as a function of the voltage. This BLS intensity at the end of the pulse is corresponding to the final magnon intensity in the frequency range accessible with BLS, which is the low frequency region of the spectrum. The ``+''-sign data points correspond to the direction in which the STT effect injects magnons, the ``$-$''-sign data points correspond to the case when the STT effect annihilates magnons (a negative voltage corresponds to a positive current density in $x$-direction and results in an injection of magnons as indicated by "+"-sign data points). We observe an increasing BLS intensity at the end of the pulse with higher voltages, indicating an increasing STT-injection. After reaching its maximum at a voltage of $U=\SI{1.45}{\volt}$ the BLS intensity decreases again, which is attributed to a decreased spin mixing conductance due to the heating \cite{Uchida.2014}, and a decreased BLS sensitivity at higher temperatures \cite{Olsson.2017}. The decreasing BLS intensity for opposite current polarity is a superposition of the annihilation of magnons due to the SHE-STT effect and the decreasing BLS sensitivity.

In the following, the effect of the changed magnon population at the end of the pulse on the threshold voltage of the BEC formation is investigated. The applied DC pulse voltage defines the temperature increase achieved by the Joule heating, and, therefore, it defines the number of excess magnons redistributed in the process of rapid cooling. Thus, the investigation of the BLS intensity as a function of the applied voltage yields threshold information. For the time evolution of the temperature, see Supplemental materials. \looseness=-1

Figure\,\ref{figure3}(c) shows the BLS intensity after the pulse is switched off, integrated in the time interval \linebreak $\tau_\mathrm{P}<t<\tau_\mathrm{P}+95$~ns. As a reference, the black curves show the integrated BLS intensities without a SHE-STT contribution ($\upmu_0 \mathbf{H}_\mathrm{ext}\parallel \mathbf{j}_\mathrm{C}$). These two curves (positive and negative current polarity) show a sudden increase at a voltage of  $U=\SI{1.42}{\volt}$, which is the threshold of the BEC formation without a SHE-STT contribution. For the case of a STT-annihilation (bright blue graph) a pronounced threshold at a higher voltage of $U=\SI{1.51}{\volt}$ is observed. \looseness=-1

In the case of STT-injection, the BLS intensity  after termination of the pulse (dark blue curve) is the interplay of the rapid cooling induced magnon redistribution and the (decaying) STT-injected magnons during the pulse. To investigate if these two contributions are a linear superposition or if the SHE-STT effect driven injection influences the redistribution process, we need to subtract the intensity originating from an exponential decay of the STT-injected magnons. This intensity can be derived as $I^\mathrm{STT}(t)=I^\mathrm{STT}_{t=\tau_{P}} \mathrm{exp}[-2(t-\tau_\mathrm{p})/\tau_\mathrm{m}]$,
where $I^\mathrm{STT}_{t=\tau_{P}}$ is the intensity at the end of the pulse [integrated value shown in Fig.~\ref{figure3}(b)], and $\tau_\mathrm{m}$ the magnon lifetime. Fitting the time evolution of the BLS intensity for the lowest voltages applied (for STT-injection, without a contribution of the rapid cooling mechanism) reveals $\tau_\mathrm{m}=\SI{34}{\nano \second}$. 
Thus, by using the starting intensity of the SHE-STT effect injected magnons we can calculate the expected magnon intensity after the pulse if no rapid cooling process would take place. 
The experimental BLS intensity after the pulse with the calculated STT-contribution subtracted is shown in Fig.~\ref{figure3}(d). For $U>\SI{1.30}{\volt}$ we find that the intensity after the end of the pulse is larger than the intensity given by the finite lifetime of the STT-injected magnons. This lowest voltage that leads to an increase of the number of magnons above the STT-injected level is identified as the threshold of the BEC formation. In summary, the shift of the BEC formation threshold  is from $U=\SI{1.42}{\volt}$ down to $\SI{1.30}{\volt}$ or up to $U=\SI{1.51}{\volt}$, corresponding to a relative change of -8\% or +6\% respectively.
This shift results from a relative SHE-STT effect driven change of excess magnon density in the order of 0.002\%, as we discuss in the Supplement. At the same time, these magnons are injected directly to the bottom of the magnon spectrum and, thus, efficiently contribute to the BEC formation. In fact, the number of SHE-STT-effect-injected magnons compares well with the density of injected magnons in parallel parametric pumping experiments, which is in the order of $10^{18}$--$10^{19}$~cm$^{-3}$. \cite{Demokritov.2006}.  \looseness=-1

The decreasing intensity at higher voltages  [$U>\SI{1.65}{\volt}$ in Fig.~\ref{figure3}(c)] requires further investigations but might be attributed to the increased temperature of the YIG. At the time, at which the magnon BEC occurs, the YIG is not yet cooled down to room temperature. Thus, at higher voltages the temperature at that time is increased, resulting in a decrease of the BLS sensitivity. The difference in the BLS intensity after the pulse between the case of $j_{\mathrm{C},x}>0$ and $j_{\mathrm{C},x}<0$ decreases with higher voltages and vanishes for the highest voltages applied. The vanishing contribution of the SHE-STT at higher voltages $U>\SI{1.8}{\volt}$ is attributed to a decreasing spin mixing conductance with higher temperatures \cite{Uchida.2014}. 

In conclusion, we found that the SHE-STT effect injects or annihilates magnons and, thus, shifts the threshold values for the rapid cooling-induced magnon BEC up to 8\%. Utilizing the SHE-STT effect opens up opportunities for the control (triggering or suppression) of the magnon BEC formation when the applied voltage is close to the threshold value.  The comparison between the magnon populations during the application of a DC pulse for the cases with and without a SHE-STT contribution has shown that the injection of magnons primarily takes place into the low energy part of the spectrum, and, thus, strongly promotes the formation of a rapid cooling induced magnon BEC. These findings suggest a new way to control the formation of a magnon BEC by the polarity of the applied current, paving the way for the utilization of macroscopic quantum states in spintronic devices.

\begin{acknowledgments} 
This research was funded by the European Research Council within the Starting Grant No. 678309 ``MangonCircuits'' and the Advanced Grant No. 694709 ``SuperMagnonics'', by the Deutsche Forschungsgemeinschaft (DFG, German Research Foundation) within the Transregional Collaborative Research Center – TRR 173 – 268565370 “Spin+X” (projects B01 and B04) and through the Project 271741898, and by the Austrian Science Fund (FWF) within the project I 4696-N.
\end{acknowledgments} 

\bibliography{output.bbl}
\newpage
\begin{center}
	\section{Supplemental Materials}
\end{center}
\subsection{BLS magnon spectra during and after the applied DC pulse}
\begin{figure}[h]
	\includegraphics[width=\textwidth]{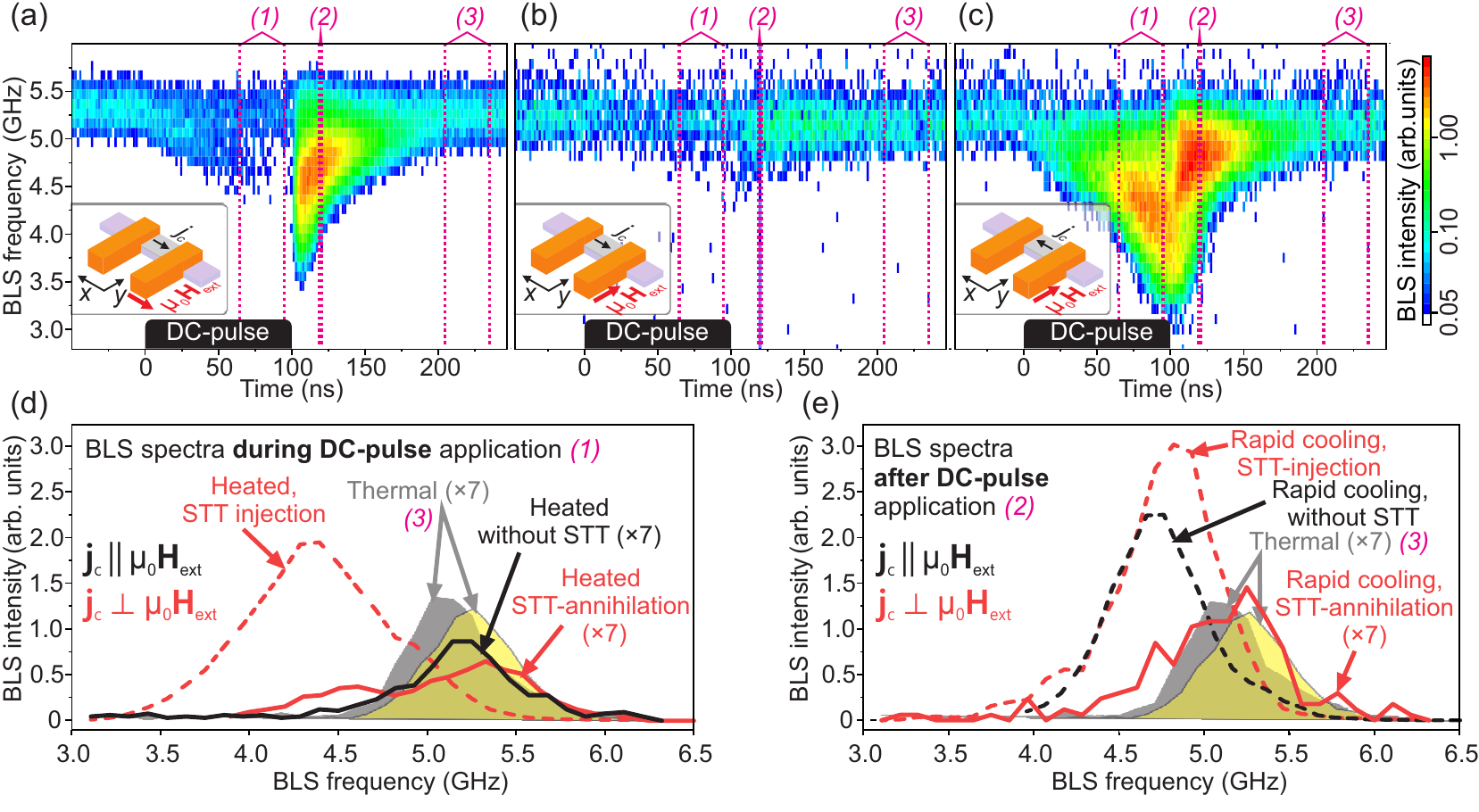}
	\caption{BLS intensity color-coded (log-scale) as a function of the BLS frequency and time. The black boxes mark the duration of the 100-ns-long heating DC pulse with an amplitude of $|U| = \SI{1.5}{\volt}$. (a) Current parallel to the external field, thus no contribution of the SHE-STT effect is expected. The achieved overpopulation after pulse termination due to the rapid cooling effect is sufficiently large to trigger the formation of a magnon BEC. (b) The current is perpendicular to the external field. The STT annihilates magnons during the pulse. The condensation of magnons is suppressed. (c) Reversed current direction in comparison to the situation in (b), the STT injects magnons. After the pulse is applied, the magnon BEC density is enhanced compared to the situation depicted in panel (a). For the exact geometries, see insets. (d) BLS spectra during the application of the DC pulse extracted in time-frame (1), as shown in (a-c). For comparison, thermal spectra were extracted in the time frame (3) for the field pointing along the short (filled grey curve) and long axis (filled yellow curve) of the waveguide. (e) BLS spectra after the DC pulse extracted in the time frame (2), as shown in (a-c), thermal spectra as in (d).}
	\label{figS1}
\end{figure}

Figures~\ref{figS1}(a-c) show the measured color-coded BLS intensities (log-scale) as a function of time and BLS frequency, as presented in Fig.~\ref{figure_2} in the main text. The amplitude of the applied DC heating pulse is $|U| = \SI{1.5}{\volt}$. 
For a better comparison of the three cases described in the main text, Figs.~\ref{figS1}(d,e) depict the extracted BLS spectra at the time frames marked in Figs.~\ref{figS1}(a-c). First, we consider injecting magnons via the STT and the consequent influence on the intensity during pulse application. The BLS intensity during the pulse application for the STT injecting magnons [dashed red line in Fig.~\ref{figS1}(d)] is more than one magnitude larger compared to the case without a SHE-STT contribution [black curve in Fig.~\ref{figS1}(d)]. Please note that the increase in intensity is observed, although the BLS sensitivity is significantly decreased at higher temperatures \cite{Olsson.2017}. Furthermore, a significant shift of the magnon frequency is observed for a SHE-STT-injection compared to the non-heated thermal spectra [filled curves in Fig.~\ref{figS1}(d), for two different directions of $\upmu_0H_\mathrm{ext}$], which is in the range of $\Delta f =\SI{1}{\giga \hertz}$. The shift is caused by the decrease in saturation magnetization due to the large number of magnons injected via the SHE-STT mechanism.
Second, in the case of a SHE-STT induced magnon annihilation [solid red line in Fig.~\ref{figS1}(d)], the BLS intensity seems to be decreased compared to the case without SHE-STT contribution [black line in Fig.~\ref{figS1}(d)]. 
Figure~\ref{figS1}(e) depicts the spectra after pulse termination during the rapid cooling process. It can be seen that the STT-annihilation of magnons suppresses the BEC formation. Here, the rapid cooling process causes only a minor increase of the BLS intensity (compare the solid red line with the thermal spectrum for the same geometry, gray-filled curve). In contrast, the SHE-STT driven injection of magnons (red dashed line) causes a larger BLS intensity than for the case of a suppressed SHE-STT contribution (black dashed line). 

\subsection{Temperature at the end of the DC pulse}
The phenomenon of magnon BEC formation by rapid cooling is driven by a rapid decrease in the phonon temperature. In the following, we address the temperature dynamics of the investigated YIG/Pt structure subjected to short heating current pulses. Therefore, we performed COMSOL-Multiphysics-based simulations, using the Electric Currents module and the Heat Transfer in Solids module included in the simulation software.

The material parameters used for the simulations are the same as given in the Supplement of Ref.~\cite{Schneider.2020}, while the corresponding temperature-resistance coefficient was adapted according to the actual experimental structure to $a=\SI{2.0883E-3}{\per \kelvin}$ and the electric conductivity was defined as $\sigma=\SI{5.1020E+6}{\siemens\per\meter}$. Both values were determined by electrical measurements combined with a Peltier element, which provided temperatures from room temperature up to \SI{350}{\kelvin}.

\begin{figure}[h]
	\includegraphics[]{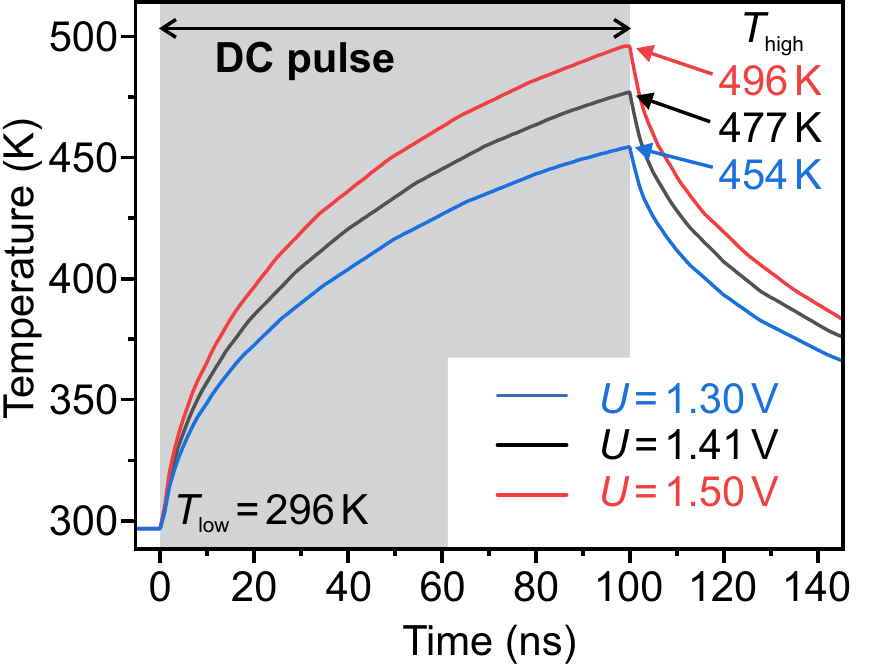}
	\caption{Temperature increase as a function of time for a 100-ns-long applied DC pulse, for three different BEC formation threshold voltages corresponding to the thresholds with additional magnon injection (blue), no SHE-STT contribution (black), and additional magnon annihilation (red). The temperatures were determined using COMSOL Multiphysics. The estimated temperature at the end of the applied pulse is indicated in the figure, and the increases are found to be of the same order as in our original experiment \cite{Schneider.2020}.}	\label{figS2}
\end{figure}

Figure~\ref{figS2} depicts the temperature as a function of time for different applied voltages in the simulation, corresponding to the threshold voltages in the experiment as indicated in the figure. The initial current densities were $j_\mathrm{C}=\SI{1.39E+12}{\ampere\per\square\meter}$, $j_\mathrm{C}=\SI{1.50E+12}{\ampere\per\square\meter}$, $j_\mathrm{C}=\SI{1.60E+12}{\ampere\per\square\meter}$. As seen from the figure, the temperature increase of $\Delta T = \SI{180}{\kelvin}$ for the BEC formation threshold voltage (without additional injection) of $U\approx\SI{1.4}{\volt}$ is in the same order as in our original study \cite{Schneider.2020}.
\newpage

\subsection{Spin injection via the SHE-STT effect and generation of excess magnons by rapid cooling}
We determined the critical voltage for the structure under investigation, exactly compensating for the magnon decay, as $U = \SI{0.95}{\volt}$ [see Fig.~\ref{figure3}(a) in the main text]. This value corresponds to an initial charge current density of $j_\mathrm{C}=\SI{1.01E+12}{\ampere\per\square\meter}$. We can describe the corresponding critical spin current density \cite{PhysRevLett.113.197203} as: 
\begin{equation}
	|j_\mathrm{s,crit}|=\tau_\mathrm{m}^{-1} d M_\mathrm{S,eff} \upgamma^{-1},
\end{equation}
where $\tau_\mathrm{m}$ is the lifetime of lowest-energy magnons, $d$ is the thickness of the magnetic layer, $M_\mathrm{S,eff}=\SI{151.2}{\kilo\ampere \per\meter}$ and $\upgamma=\SI{28}{\giga\hertz\per\tesla}$ is the gyromagnetic ratio. Hence, the spin current density, which compensates exactly for the damping of the lowest-energy magnon modes, is $|j_\mathrm{s,crit}| = \SI{5.41E-6}{\kilo\gram\per\square\second}$ or, in units of integer spin, \linebreak $|j_\mathrm{s,crit}^*|=\SI{5.13E+28}{\per\square\meter\per\second}\hbar$. Hence, using the simplest definition of a current $j_\mathrm{s,crit}^* /\hbar =N /(\Delta t A)$, we find that we inject a total number of $N_\mathrm{inj} = 2.05\times 10^{10}$ spins for an applied critical voltage of $U = \SI{0.95}{\volt}$. Here, we considered a pulse duration of $\Delta t =\tau_\mathrm{P}=\SI{100}{\nano\second}$ and an interface area of $A = \SI{2}{\micro\meter} \times \SI{2}{\micro\meter}$. In good approximation, each of these spin moments transferred into the YIG spin system results in the excitation of one magnon. Finally, assuming a linear dependence on the applied voltage, we can express the total number of injected magnons for an applied voltage $U$ in our experiment as $N_\mathrm{inj}(U) = \SI{2.16E+10}{\per \volt} U$, corresponding to an injected quasiparticle density of $n_\mathrm{inj} (U) = \SI{2.16E+16}{\per \cubic\centi\meter \per \volt}U$. 

Considering the voltages applied in the experiment, we find that this value (e.g., $n_\mathrm{inj} = \SI{2.81E+16}{\per \cubic\centi\meter}$ for $U = \SI{1.3}{\volt})$ is around two orders of magnitude smaller than typical values of particle densities injected in BEC experiments with parametric pumping, being on the order of $10^{18}$~cm$^{-3}$ \cite{Demokritov.2006}. Nevertheless, it should be noted that the STT effect only slightly changes the BEC threshold in our experiments rather than inducing BEC formation like in the case of the parametric pumping experiments. Moreover, the STT effect injects magnons primarily to the lowest part of the magnon spectrum. In contrast, the magnons injected by the parametric pumping process typically have higher energies, and a thermalization process is required before they reach the bottom of the spectrum. 
Furthermore, accounting for the relaxation of magnons with a characteristic lifetime of $\tau_\mathrm{m}=\SI{34}{\nano\second}$, we can calculate the number of generated excess magnons at the end of the pulse as 
\begin{equation}
	n_\mathrm{inj}^{t=\tau_{\mathrm{P}}} (U)=\int_{0}^{\tau_\mathrm{P}} \frac{n_\mathrm{inj}(U)}{\tau_\mathrm{P}}\exp[{-(\tau_{\mathrm{P}}-t})/\tau_{\mathrm{m}}] \mathrm{d}t,
\end{equation}
giving a SHE-STT driven excess magnon density of $n_\mathrm{inj}^{\SI{100}{\nano\second}}(U=\SI{1.3}{\volt})=\SI{9.04E+15}{\per\cubic\centi\meter} $ at the end of the pulse. 

We can further compare this value of SHE-STT-driven excess magnons to the number of excess magnons generated by the rapid cooling mechanism. To calculate the number of excess magnons generated by the rapid cooling mechanism, we can approximate the excess magnon density induced for a temperature drop from $T_\mathrm{high} = \SI{454}{\kelvin}$ to $T_\mathrm{low} = \SI{296}{\kelvin}$ ($U = \SI{1.3}{\volt}$) via:
\begin{equation}
	\frac{n_\mathrm{BE}(\upmu=0, T= \SI{454}{\kelvin})}{n_\mathrm{BE}(\upmu=0, T= \SI{296}{\kelvin})}\approx 1.6,
\end{equation}
where
 \begin{equation}
 	n_{\mathrm{BE}}(\upmu,T)=\int_{E_\mathrm{min}}^{E_\mathrm{max}}\frac{D(E)}{\exp[{\frac{E-\upmu}{k_\mathrm{B}T}}]-1}\mathrm{d}E,
 \end{equation}
with the density of states $D(E)$. Here, $E_\mathrm{min}/h=\SI{5}{\giga\hertz}$ is the bottom of the magnon spectrum and $E_\mathrm{max}/h=\SI{7}{\tera\hertz}$ is the edge of the first Brillouin zone in YIG. Thus, in the process of rapid cooling, for instance, for the case of $U = \SI{1.3}{\volt}$, an excess magnon density on the order of the thermal magnon density is generated, hence, of \linebreak $n_\mathrm{RC}=(\frac{n_\mathrm{BE}(\upmu=0, T= \SI{454}{\kelvin})}{n_\mathrm{BE}(\upmu=0, T= \SI{296}{\kelvin})}-1) n_\mathrm{room}\approx \SI{6E+20}{\per\cubic\centi\meter}$, assuming that the density of magnons at room temperature is $n_\mathrm{room}\approx 10^{21}$~cm$^{-3}$ \cite{Demokritov.2006}. However, due to their thermal origin, these magnons are distributed over the whole spectral range reaching up to several THz for YIG. Thus, the high-energy magnons decay fast and, thus, weakly contribute to the BEC formation, as shown in the original paper \cite{Schneider.2020}. 

Hence, recapitulating the case of $U = \SI{1.3}{\volt}$ in our experiment, this simple estimation shows that approximately 0.002 \% variation of the number of excess magnons in the bottom part of the magnon spectrum results in a BEC formation threshold shift of about 6 \% to 8 \%. The reason for this comparably large effect is that the additionally SHE-STT-injected magnons have low energy and, thus, contribute more significantly to the BEC formation. 
\newpage
\subsection{BLS spectra for different geometries as a function of the applied voltage}

We discussed the shift of the threshold of a rapid-cooling-induced magnon BEC formation based on the annihilation/injection of magnons via the SHE-STT effect during the pulse. For a slightly supercritical voltage range, we observe an injection of magnons to the bottom of the magnon spectrum in a rather broad frequency range, promoting the subsequent redistribution of the rapid cooling magnon excess to the bottom (see Fig.~\ref{figure_2} in the main manuscript, and Fig.~\ref{figS1}). However, the change of the external field direction changes the magnon mode profile in the dipolar regime. Furthermore, for large enough voltages and sufficiently long DC pulses, the SHE-STT-effect-based injection of magnons is known to trigger the formation of auto-oscillations, which can affect the spectral magnon distribution during the pulse. Both effects, the excitation of the auto-oscillations and the modified mode profile, might have consequences for the resulting redistribution of magnons, and therefore they are addressed in more detail here.

\begin{figure}
	\includegraphics{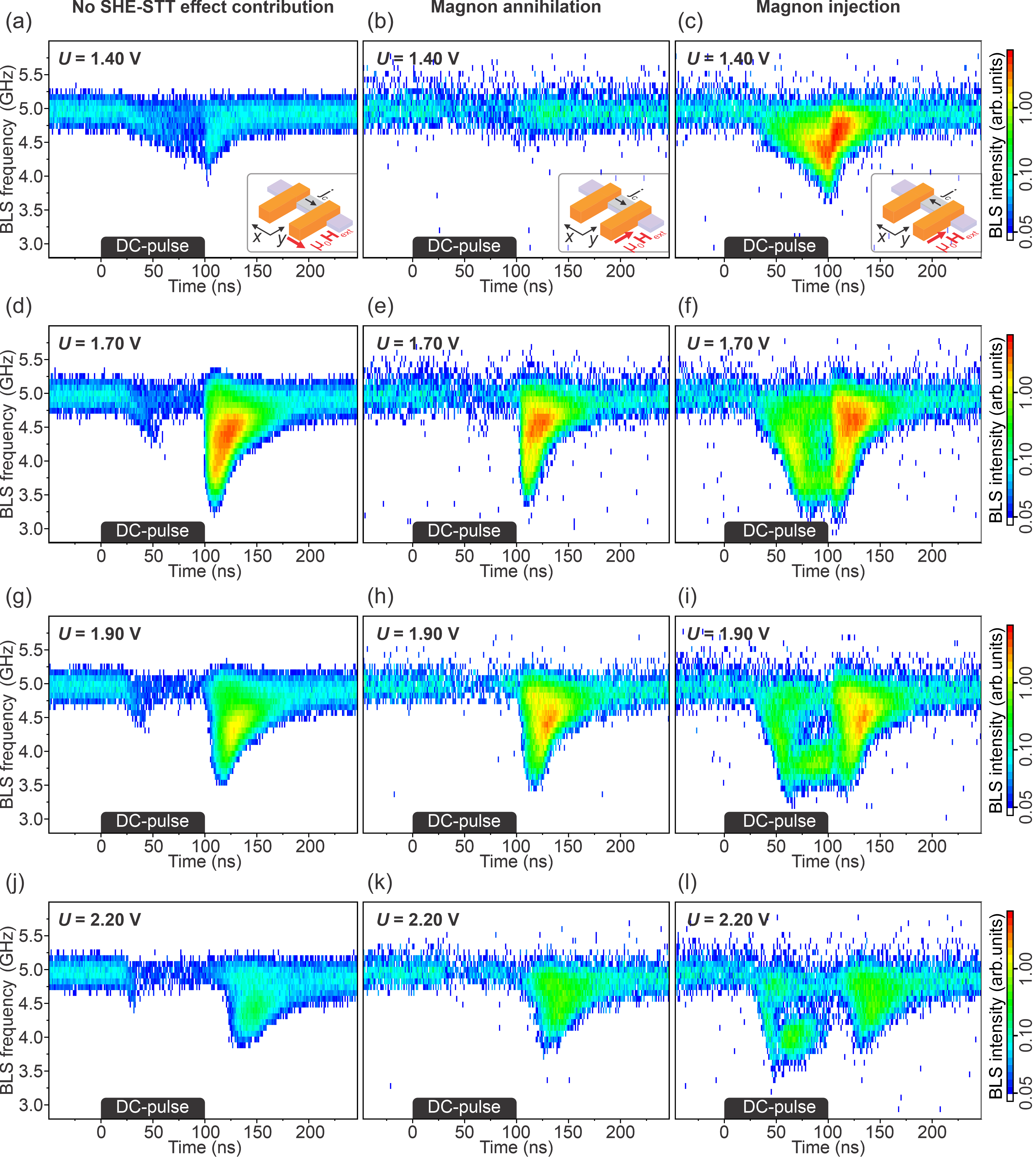}
	\caption{BLS intensities and frequencies as a function of time for different applied voltages and for the three geometries investigated. First column: current and external field are parallel, no contribution of the SHE-STT effect, second column: current and field are aligned perpendicularly, magnon annihilation via the SHE-STT effect during the pulse, third column: current and field are aligned perpendicularly, magnon injection via the SHE-STT effect during the pulse. Insets in (a-c) depict the corresponding geometries. (a-c) $U=\SI{1.40}{\volt}$, only for the case of magnon injection during the pulse, the threshold of rapid cooling induced BEC formation is reached (c), and an increased intensity after pulse termination is observed. (d-f) $U=\SI{1.70}{\volt}$, increased intensity after pulse termination for all geometries. For magnon injection (f), a lower–frequency mode is populated during the pulse, attributed to the excitation of a bullet mode. Rapid-cooling-induced condensation takes place partially into the bullet mode state. (g-i) $U=\SI{1.90}{\volt}$, analogous to (d-f) with slightly decreased intensities after pulse termination, attributed to an overheating of the magnon system. The additionally excited bullet mode in (i) has a minor influence on the spectrum after pulse termination. (j-l) $U=\SI{2.20}{\volt}$, SHE-STT effect injection decreases during the pulse (l) attributed to the higher temperature reached. Consequently, similar spectra after pulse termination are observed for the different geometries.}
	\label{figS3}
\end{figure}

Figure~\ref{figS3} shows the BLS spectra as a function of time for the three different geometries investigated and various applied voltages (see insets for the geometry used in the experiment). As can be seen, we observe similar spectra after pulse termination just above the threshold of magnon BEC formation [compare Figs.~\ref{figS3}(c), (d) and (e)]. The main difference observed is the slightly higher frequency of the redistributed magnons for the case of magnon injection during the pulse [Fig.~\ref{figS3}(c)] compared to the other cases of slightly supercritical voltages (without any SHE-STT effect contribution or for annihilation of magnons during pulse [Figs.~\ref{figS3}(d,e)]). The reason for this is the lower temperature reached at the end of the pulse in the first case, which results in a less pronounced shift of the magnon dispersion. Apart from that, we do not observe a significant influence of the changed field direction, as seen from comparing the spectra depicted in Figs.~\ref{figS3}(d,e) and~\ref{figS3}(g,h).

For larger voltages applied to inject magnons, we observe the SHE-STT-effect-driven excitation of the magnon bullet mode. The reason for the excitation of the magnon bullet, which arises from nonlinear effects, is the reduced effective magnon relaxation rate of this particular mode. The low effective relaxation rate results from the bullet mode’s low frequency and self-localization, preventing propagation losses.

As shown in Fig.~\ref{figS3}(f), we observe the bullet mode formation at sufficiently high voltages for the magnon injection case (starting at $U\approx\SI{1.55}{\volt}$). This bullet formation process changes the spectral magnon distribution during the pulse with respect to the cases of magnon annihilation during pulse or without a SHE-STT effect contribution. Consequently, the spectra of redistributed magnons change after pulse termination. In particular, we find that the magnons are partially redistributed to this bullet mode state  at $f \approx \SI{3.7}{\giga\hertz}$. The effect of the established auto-oscillation regime on the rapid-cooling-induced Bose--Einstein condensation of magnons is out of the scope of this work, and the results will be published elsewhere. However, for larger volt-ages applied, this effect vanishes again [see Figs.~\ref{figS3}(i,l)], which is attributed to the fact that the magnon injection during the current pulse becomes less efficient with increasing temperature and the increase of the temperature of the magnon gas.

	\end{document}